\documentclass[preprint2]{aastex}

\newcommand{\Msun}{\ensuremath{\mathrm{M}_\odot}}

\shorttitle{Can very massive Population III stars produce a super-collapsar?}
\shortauthors{Yoon et al.}

\begin{document}


\title{Can very massive Population III stars produce a super-collapsar?}


\author{Sung-Chul Yoon\altaffilmark{1}, Jisu Kang\altaffilmark{1}, and Alexandra Kozyreva\altaffilmark{2}}
\altaffiltext{1}{Department of Physics and Astronomy, Seoul National University, Gwanak-gu, Gwanak-ro1, 151-742, Seoul, South Korea} 
\email{yoon@astro.snu.ac.kr}
\altaffiltext{2}{Argelander Institute for Astronomy, University of Bonn, Auf dem H\"ugel 71, D-53121, Bonn, Germany}




%

\begin{abstract}
A fraction of the first generation of stars in the early Universe
may be  very massive ($\gtrsim 300~\mathrm{M_\odot}$) as they form in metal-free
environments. Formation of black holes from these stars can be accompanied by
supermassive collapsars to produce long gamma-ray bursts of a unique type
having a very high total energy ($\sim 10^{54}~\mathrm{erg}$) as recently
suggested by several authors.  We present new stellar evolution models of very
massive Population III stars including the effect of rotation to provide
theoretical constraints on super-collapsar progenitors.  We find that the
angular momentum condition for super-collapsar can be fulfilled if magnetic
torques are ignored, in which case Eddington-Sweet circulations play the
dominant role for the transport of angular momentum.  We further find that the
initial mass range for super-collapsar progenitors would be limited to
$300~\mathrm{M_\odot} \lesssim M \lesssim 700~\mathrm{M_\odot}$.  However, all of our very
massive star models of this mass range end their lives as red supergiants rather than blue
supergiants, in good agreement with most of the previous studies.   
The predicted final fate of these stars is either a jet-powered 
type IIP supernova or an ultra-long, relatively faint gamma-ray transient, 
depending on the initial amount of angular momentum. 
\end{abstract}
\keywords{stars: evolution --- stars: rotation --- stars: Population III --- gamma-ray bursts --- early universe}

\section{Introduction}

Population III stars are the main sources of ionizing photons and the only
producers of heavy elements in the early Universe.  Their exact role on the
reionization history and the chemical evolution largely depends on their
initial mass functions.  In the metal-free environment where the first stars
form, thermal cooling of star-forming regions is dominated by hydrogen
molecules, which are much less efficient coolants than heavy elements and
dusts. It is therefore widely believed that Population III (Pop III) stars are
systematically more massive than Pop II and Pop I stars~\citep[e.g.,][]{Abel02,
Bromm02}. 

According to recent numerical simulations, most of Pop III stars would have a
mass range of $10 - 1000~\mathrm{M_\odot}$~\citep[e.g.,][]{Omukai03, Ohkubo09,
Bromm09, Hosokawa11, Stacy12, Hirano14}, while formation of more massive Pop
III stars ($ > 1000~\mathrm{M_\odot}$) still remains a good possibility
\citep{Hosokawa12}.  A significant fraction of them would produce 
core-collapse supernovae for initial masses of $10 - 100~\mathrm{M_\odot}$, and
pulsational pair-instability or pair-instability supernovae for  $100 -
300~\mathrm{M_\odot}$ \citep{Heger02, Umeda02}.  
According to the recent numerical simulations, the first stars
may rotate at a velocity close to the critical rotation~\citep{Stacy11, Stacy13}, and
these mass ranges can be
significantly lowered if rapid rotation induces 
strong chemical mixing inside stars~(\citealt{Yoon12, Chatzopoulos12}; cf. \citealt{Glatzel85}).
Gamma-ray bursts may also occur with sufficiently high initial angular
momentum, for about $12 - 80~\mathrm{M_\odot}$~\citep{Yoon12}.  
Detection of these events at $z \simeq 20$ with next generations of telescopes
will give us invaluable information on the nature of Pop III
stars~\citep[e.g.,][]{Kawai06, Greiner09, Salvaterra09, Tanvir09, Cucchiara11}.

On the other hand, Pop III stars with initial masses of $300 - 50000$~\Msun{}
are supposed to directly collapse to a black hole \citep[e.g.,][]{Fryer01,
Heger02}.  In particular,  those with $M_\mathrm{init} > 1000~\mathrm{M_\odot}$
are often considered seeds for super-massive black holes that are found at high
redshift~\citep[e.g.][]{Madau01, Schneider02}. Recently several authors
suggested that formation of black holes in very massive stars can be
accompanied by a super-collapsar (i.e, collapsar with a black hole mass higher
than a few hundreds solar masses) to produce relativistic
jets~\citep{Komissarov10, Meszaros10, Suwa11, Maio14}.  Given that the
considered black holes are very massive compared to the case of ordinary
gamma-ray bursts, there might be unique observational signatures from these
super-collapars, depending on the details of the progenitor structure.  For
example, the amount of the total energy released by jets could be significantly
larger than those of ordinary gamma-ray bursts \citep{Komissarov10,
Meszaros10}.

One of the key conditions for having a collapsar event  is rapid rotation of
the infalling matter onto the black hole, of which the specific angular
momentum must be high enough to form an accretion disk \citep[i.e., $j \gtrsim
10^{16} - 10^{19}~\mathrm{cm^2~s}$ for the black hole mass of 3 to
1000~$\mathrm{M_\odot}$;][]{Woosley93, MacFadyen99}: If the black hole mass is
higher than a few hundreds solar masses, the corresponding radius of the disk
is  too large for neutrino annihilation to produce a relativistic jet
\citep{MacFadyen99}.  Therefore, another key ingredient for super-collapsar is
strong magnetic fields (i.e., $B\gtrsim 10^{13}$ G around the black hole)
such that the rotational energy of the black hole may be extracted by the
Blanford-Znajek mechanism \citep{Komissarov10, Meszaros10}. The progenitors of
super-collapsars should be both rapidly rotating and strongly magnetic.

\citet[][hearafter, YDL12]{Yoon12} calculated evolutionary models of Pop III
stars including rotation and magnetic fields. In their models the
Tayler-Spruit (TS) dynamo \citep{Spruit02} plays a key role in transporting
angular momentum inside stars.  Because of the efficient angular momentum
transport from the convective core to the outer layers via magnetic torques,
stars can maintain near-rigid rotation throughout the main sequence, and lose
angular momentum via mass shedding as the surface reaches critical rotation. 
This effect becomes more important for a very massive Pop III
star of which the surface luminosity is close to the Eddington limit because
it can reach the critical rotation even with a fairly low rotation velocity.
As a result, Pop III stars more massive than about 300~\Msun{} cannot retain
enough angular momentum to produce a super-collapsar, according to YDL12.

The TS dynamo adopted by YDL12 is one of the most considered explanations for
the rapid transport of angular momentum in radiative layers of stars, which is
implied by a number of observations, in particular with low-mass stars
\citep[e.g.,][]{Eggenberger05,  Suijs08, Eggenberger12, Cantiello14}.
However, the efficiency of the TS dynamo is still a matter of debate in the
community \citep{Braithwaite06, Zahn07, Denissenkov10,  Arlt11}. The strong
braking implied in low-mass stars might be instead related to some other
mechanisms like internal gravity waves \citep{Zahn97, Talon02}, which have not
been well understood in the context of massive stars and thus might not be
necessarily relevant.  If the transport of angular momentum in
massive stars is much less efficient than what is predicted with the TS dynamo,
they could retain a large amount of angular momentum inside the core, while
magnetic fields of large scales might be generated in convective layers before
collapse or in the accretion disk after black hole formation, thus fulfilling
the necessary conditions for super-collapsar.  Given the importance of
super-collapsar events as a probe on the nature of Pop III stars, it is worthwhile to
investigate this possibility in some detail to provide a theoretical boundary condition
for super-collapsar progenitors.   For this purpose, we present new
evolutionary models of very massive Pop III stars ($M_\mathrm{init} \gtrsim
300~\mathrm{M_\odot}$) without magnetic torques according to the TS dynamo.
This means that we only consider Eddington-Sweet circulations and other
hydrodynamic instabilities for the transport of angular momentum. 

We briefly explain the numerical methods used for our models in
Sect.~\ref{sect:method}.  In the section that follows (Sect.~\ref{sect:zams}),
we describe the properties of very massive Pop III stars on the zero-age main
sequence (ZAMS). In particular, we show that the angular momentum condition for
collapsar becomes more difficult to meet for a higher initial mass because the
Eddington factor at the stellar surface becomes larger.  In
Sect.~\ref{sect:evol}, we present the evolutionary models including rotation
and discuss the initial conditions of Pop III stars
needed for super-collapsar progenitors.  We discuss the implications of our
results for the final fate of super-collapsar progenitors and the relevant
uncertainties of our models in Sect.~\ref{sect:implications}.  We conclude our
discussions in Sect.~\ref{sect:conclusions}.

\section{Numerical methods}\label{sect:method}

We used the one-dimensional hydrodynamic stellar evolution code that is
described in YDL12 and \citet{Kozyreva14a}.  This code implicitly solves
the stellar structure equations for which the Henyey-type method is adopted.
The OPAL opacity table~\citep{Iglesias96} is used for $T> 10^4$~K and
the prescription by \citet{Alexander94} for lower temperatures.  We follow
\citet{Endal76} to consider the effect of the centrifugal force on the stellar
scructure, and \citet{Heger00} and \citet{Heger05} for the redistribution of
angular momentum, which is approximated as a diffusive process. We used the
same physical parameters for convection, semi-convection and
rotationally-induced chemical mixing and transport of angular momentum, but the
TS dynamo is not included in the present study except for one model sequence.
This means that the considered transport processes due to rotation include
Eddington-Sweet circulations, the secular shear instability and the Goldreich,
Schubert and Fricke (GSF) instability as explained in \citet{Heger00}.  The
overshooting is applied for 0.335 times local pressure scale heights above the
convectively unstable core and the adopted semi-convection parameter
($\alpha_\mathrm{semi}$; \citealt{Langer83}) is 1.0, following \citet{Brott11}.
Compared to YDL12, the nuclear network was improved to treat silicon burning as
discussed in \citet{Kozyreva14a} and  \citet{Kozyreva14b}. In some of our
models, the silicon burning was followed with 13 main isotopes. In this way the
main alpha-chain  can be described without considering neutronization that does
not play an important role in the structure of very massive stars undergoing
the pair-instability.

The treatment of mass loss for Pop III stars and its physical justification are
fully described in YDL12.  
In short, the mass loss in our models is dominated
by the centrifugally-driven mass shedding, which is treated as the following:
\begin{equation}\label{eq1}  
\dot{M}(v) =
\mathrm{max}\left[10^{-14}~M_\odot~\mathrm{yr^{-1}}, \dot{M}(v=0)\right]
\left(\frac{1}{1-\Omega}\right)^{0.43}~, 
\end{equation} 
where
\begin{equation}\label{eq2}  
\Omega :=
\frac{v}{v_\mathrm{crit}}~~\mathrm{and}~~v_\mathrm{crit} =
\sqrt{\frac{GM}{R}(1-\Gamma)}~.  
\end{equation} 
The critical rotation $v_\mathrm{crit}$ represents the modified Keplerian limit
considering the effect of radiation when the stars approach the Eddington
limit,  for which  the Eddington factor $\Gamma$ is given by $\kappa L/(4\pi GMc)$. 
$\dot{M}(v=0)$ is the wind mass loss rate for the non-rotating case, for which
we followed \citet{Kudritzki89} and \citet{Nieuwenhuijzen90} 
for $T_\mathrm{eff} > 10^4$ and $T_\mathrm{eff} \le 10^4$, respectively, 
with metallicity dependence of $Z^{0.69}$. Here, the metallicity 
means the total mass fraction of CNO elements at the surface, 
and in this way the enhancement of the mass loss rate resulting from surface enrichment of heavy elements by chemical mixing
is taken into account (See YDL12 for more discussion on this issue).

\section{Properties on the zero-age main-sequence}\label{sect:zams}

Before presenting the evolutionary models, here we discuss the properties of
very massive Pop III stars on the ZAMS.  In Table~1, we provide the information
on the ZAMS models for the mass range of 300 -- 20000~\Msun{}.  Here we assume
that stars on the ZAMS rotate as a solid body, which can be justified
because of the expected rapid transport of angular momentum in chemically
homogeneous stars by convection and Eddington-Sweet
circulations~\citep[e.g.,][see also the discussion below]{Haemmerle13}. 

The surface luminosities of these very massive stars are
close to the Eddington limit: the Eddington factor varies from $\Gamma = 0.774$
at 300~\Msun{} to $\Gamma = 0.996$ at 20000~\Msun{}.  In consequence, these
stars cannot have a very high critical rotation velocity, and the total amounts
of angular momentum should be limited accordingly.   The corresponding
$v_\mathrm{crit}/v_\mathrm{K}$ ($=\sqrt{1-\Gamma}$; see Eq.~\ref{eq2}) with
solid-body rotation changes from 0.48 to 0.06.  Because the stellar radius is
larger with higher mass, both the total angular momentum and the specific
angular momentum at the critical rotation gradually increase with increasing
mass up to about 10000~\Msun{}, despite the fact that
$v_\mathrm{crit}/v_\mathrm{K}$  decreases.  However,
$v_\mathrm{crit}/v_\mathrm{K}$ becomes so small for 20000~\Msun{} that the
specific angular momentum becomes smaller than that of 10000~\Msun{} star.
Another factor that limits the amount of angular momentum is the change of
stellar structure as $\Gamma$ approaches zero. The  dimensionless radius of
gyration (i.e., $k = J/(v_\mathrm{rot}MR)$; see Table~1) decreases with
increasing mass and therefore the amount of angular momentum for a given set of
$v_\mathrm{crit}$, $M$ and $R$ should decrease as well.  

\begin{figure*}
\epsscale{1.00}
\plotone{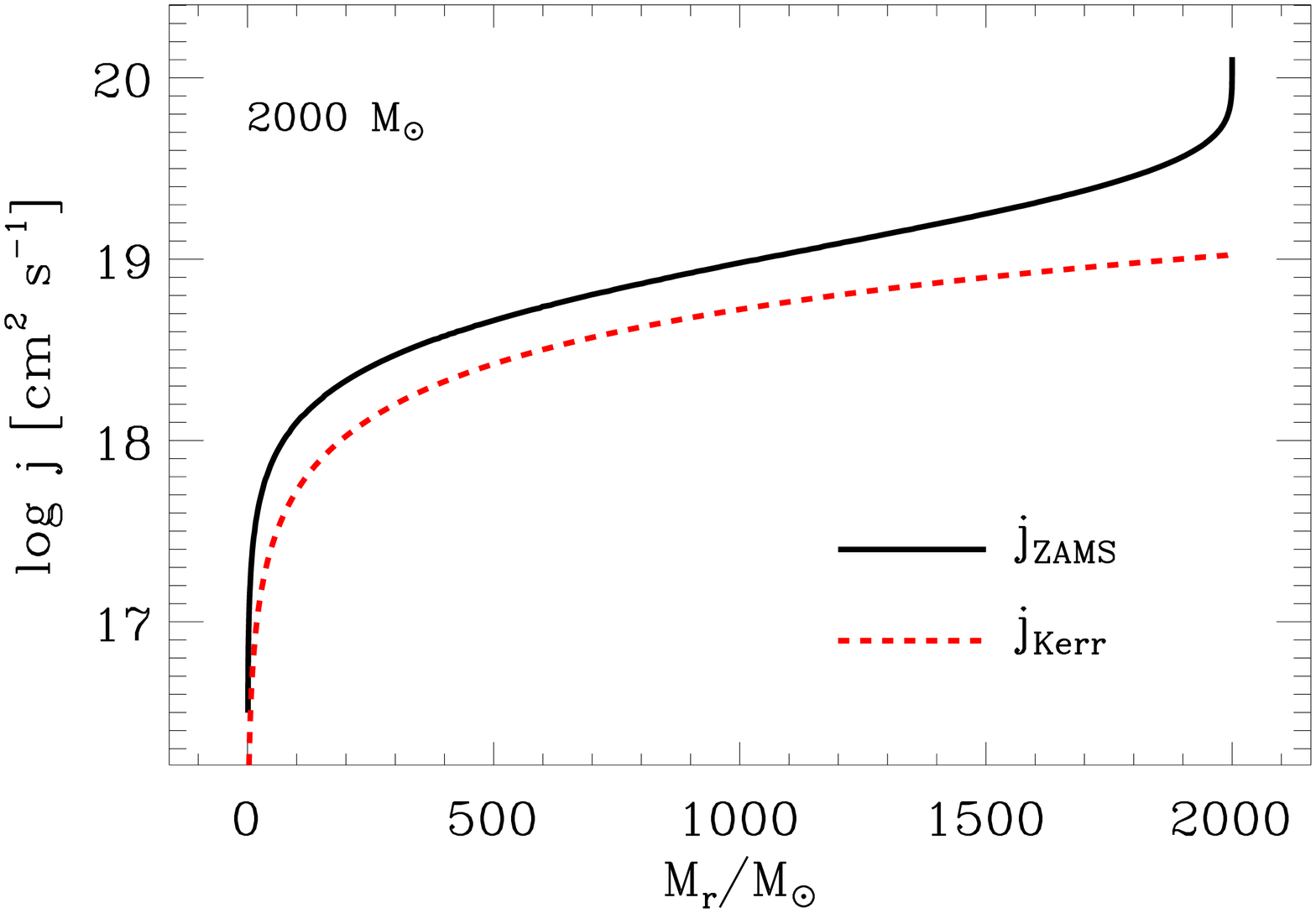}
\plotone{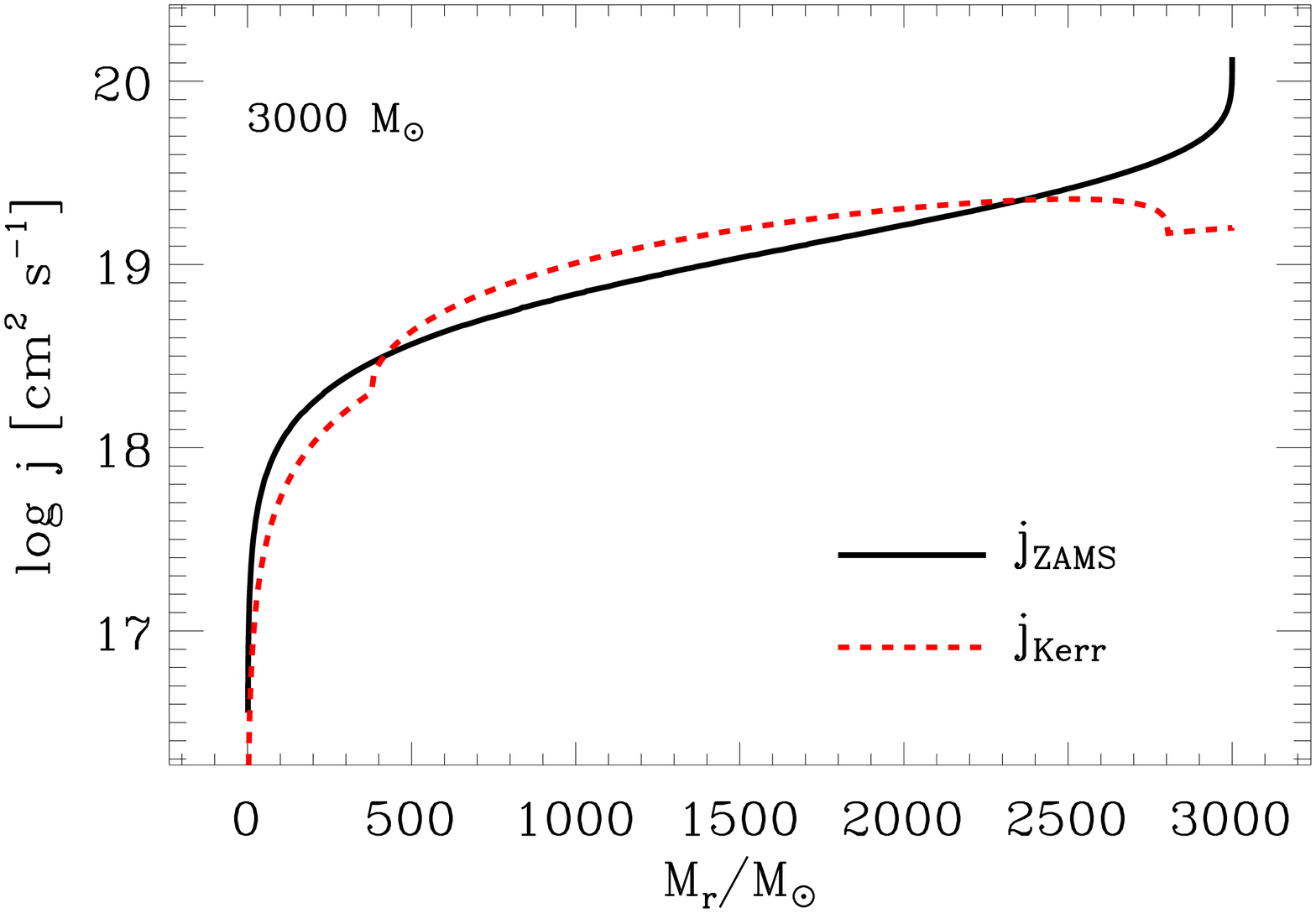}
\plotone{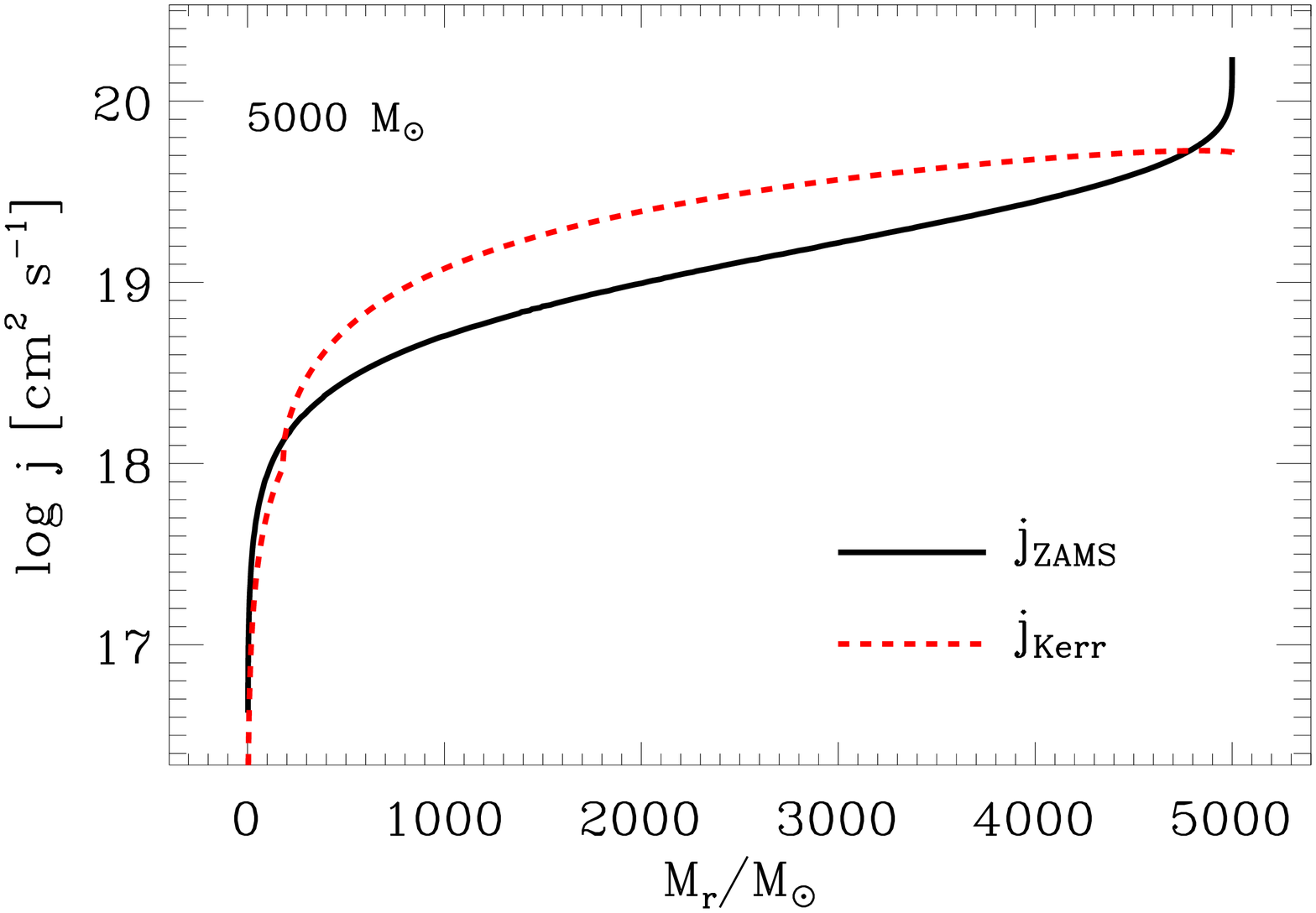}
\plotone{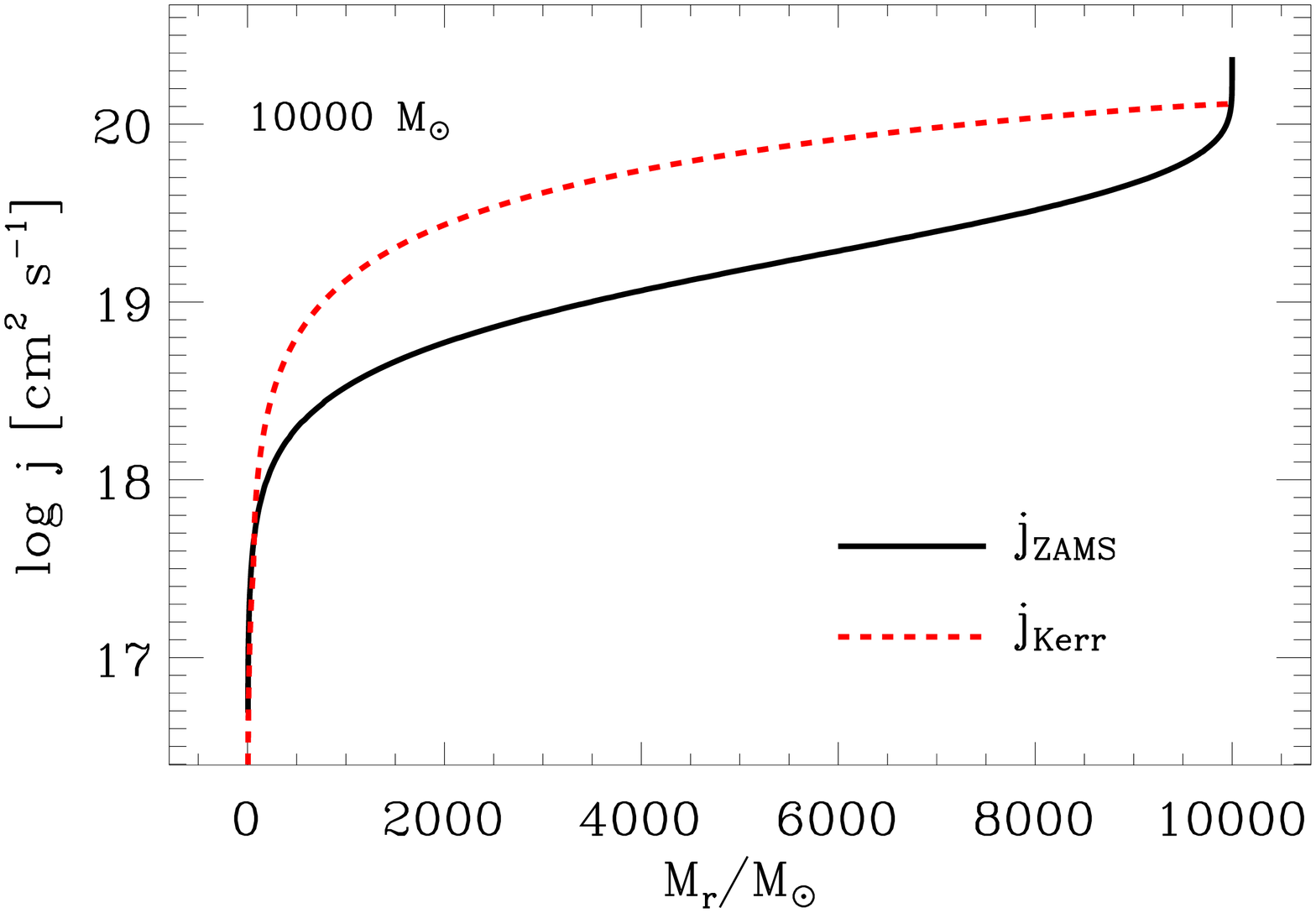}
\caption{Specific angular momentum distribution in the Pop III models on the ZAMS for 2000, 3000, 5000, and 10000~\Msun{}, 
as a function of the mass-coordinate (solid line), compared to the specific angular momentum at the last stable orbit
around the Kerr black hole that would form with the given mass and angular momentum (dashed line). 
}\label{fig:jzams}
\end{figure*}

This property leads to the following important conclusion: Pop III stars with
$M_\mathrm{i} > 3000$~\Msun{} at their birth cannot contain enough angular
momentum in their cores to produce a collapsar because of the effect of high
radiation pressure,  as illustrated in Fig.~\ref{fig:jzams}. Only the outermost
layers have specific angular momentum above the critical limit in these stars,
and therefore production of an energetic gamma-ray burst cannot occur (see
discussion below).  This fact is based only on the structure of Pop III stars
on the ZAMS, and does not depend on the uncertain physical processes like mass
loss that has strong impact on the stellar evolution.  This gives a strong
upper mass limit for super-collapsar progenitors.  Given that $M_\mathrm{i}
\gtrsim 300$~\Msun{} is required to avoid a pair-instability explosion, only
Pop III stars in the mass range of $300~\Msun{} \lesssim M_\mathrm{i} \lesssim
3000$~\Msun{} can be considered potential super-collapsar progenitors that can
produce an energetic gamma-ray burst, in terms of the initial angular momentum
budget.  Therefore,  this result effectively rules out the possibility of an
energetic gamma-ray burst with formation of intermediate-mass black holes of
$\sim 10^4~\mathrm{M_\odot}$ from super-massive Pop III stars that can be seeds
for super-massive black holes in high-redshift quasars \citep[cf.][]{Spolyar09,
Freese10}.

\begin{deluxetable}{cccccccc}
\tabletypesize{\scriptsize}
\tablecaption{Physical properties of the ZAMS models}\label{tab1}
\tablewidth{0pt}
\tablehead{
\colhead{$M_\mathrm{i}$} & 
\colhead{$T_\mathrm{eff}$} & 
\colhead{$\log L/L_\odot$} & 
\colhead{$\Gamma$} & 
\colhead{$v_\mathrm{crit}/v_\mathrm{K}$} & 
\colhead{$\log J$} & 
\colhead{$j$} & 
\colhead{$k$} 
}
\startdata
[$\mathrm{M_\odot}$] & [$10^3$~K] &     &        %
 &  & [$\mathrm{erg~s}$] & [$\mathrm{10^{19}~cm^2~s^{-1}}$] &  \\   
\hline
300 & 106.6   & 6.8 & 0.774  & 0.48 &  54.74 & 0.93  & 0.134  \\
500 & 108.9   & 7.1 & 0.836  & 0.40 &  55.05  & 1.13  & 0.130\\
1000 &110.7   & 7.5 & 0.885  & 0.32 &  55.46  & 1.46  & 0.122\\
2000 &111.3   & 7.8 & 0.927  & 0.25 &  55.86  & 1.83  & 0.114\\
3000 & 111.3  & 8.0 & 0.954  & 0.22 &  56.08  & 2.04  & 0.110 \\
4000 & 111.2  & 8.1 & 0.964  & 0.19 &  56.24  & 2.19  & 0.107\\
5000 & 111.1  & 8.2 & 0.970  & 0.17 &  56.36  & 2.31  & 0.104 \\
7000 & 110.7  & 8.4 & 0.978  & 0.15 &  56.54  & 2.47  & 0.100\\
10000 & 110.3 & 8.5 & 0.986  & 0.12 &  56.70  & 2.50  & 0.096 \\
20000 & 108.4 & 8.9 & 0.996  & 0.06 &  56.88  & 1.92  & 0.085\\
\enddata
\tablecomments{Each column has the following meaning:
$M_\mathrm{i}$: initial mass, 
$T_\mathrm{eff}$: effective temperature in units of $10^3$ K, 
$\log L/L_\odot$: surface luminosity in units of the solar luminosity, 
$\Gamma$: Eddington factor at the surface,
$v_\mathrm{crit}/v_\mathrm{K}$: the critical rotation for the given $\Gamma$ at the equatorial surface in units of the Keplerian value with rigid rotation, 
$\log J$: the total angular momentum at the critical rotation,
$j$: the specific angular momentum at the critical rotation, 
$k$: the dimensionless radius of gyration at the critical rotation (i.e., $J = kv_\mathrm{crit}MR$). 
}
\end{deluxetable}

\section{Evolutionary models}\label{sect:evol}

\begin{deluxetable}{ccccccccccccc}
\tabletypesize{\scriptsize}
\tablecaption{Physical properties of the evolutionary models}\label{tab2}
\tablewidth{0pt}
\tablehead{
\colhead{Seq. No.} & 
\colhead{$M_\mathrm{i}$} & 
\colhead{$v_\mathrm{i}$} & 
\colhead{$v_\mathrm{i}/v_\mathrm{c}$} & 
\colhead{$\log J_\mathrm{i}$} & 
\colhead{$T_\mathrm{c, f}$} & 
\colhead{$M_\mathrm{f}$} & 
\colhead{$T_\mathrm{eff,f}$} & 
\colhead{$R_\mathrm{f}$} & 
\colhead{$\log J_\mathrm{f}$} &
\colhead{$M_\mathrm{He}$} &
\colhead{$a_\mathrm{He}$} 
}
\startdata
      & [$\mathrm{M_\odot}$] & [$\mathrm{km~s^{-1}}$] &   &  [$\mathrm{erg~s}$] &     [$10^9 \mathrm{K}$] &  %
 [$\mathrm{M_\odot}$] &  [K]   & [$\mathrm{R_\odot}$] &[$\mathrm{erg~s}$] & [$\mathrm{M_\odot}$] &    \\   
\hline
S300A &   300.0 &   274.0 &     0.2 &    53.9 &     1.0 &   298.5 &  4380 & 5563.7 &    53.7 &   153.8 &     0.6  \\
S300B  &   300.0 &   536.8 &     0.4 &    54.2 &     1.0 &   291.5 & 4081 & 5179.2 &    53.8 &   158.3 &     0.9 \\
S300C  &   300.0 &  1276.2 &     0.9 &    54.6 &     5.4 &   264.0 & 4271 & 4769.8 &    53.8 &   200.9 &     1.0  \\
S300M  &  300.0 &  1276.2 &     0.9 &    54.6 &     0.6 &   258.1 &  4442 & 5642.9 &    53.2 &   189.7 &     0.1  \\
S500A &   500.0 &   454.5 &     0.4 &    54.4 &     0.9 &   493.0 &  4458 & 7530.7 &    54.3 &   254.4 &     0.6 \\
S500B &   500.0 &   598.6 &     0.5 &    54.6 &     0.9 &   492.1 &  4348 & 7912.7 &    54.3 &   258.6 &     0.9 \\
S500C &   500.0 &  1110.8 &     0.8 &    54.9 &     6.9 &   464.9 &  4157 & 7022.7 &    54.4 &   276.8 &     1.0 \\
S700A &   700.0 &  1110.6 &     0.8 &    55.1 &     0.9 &   627.9 &  4508 & 8650.9 &    54.6 &   417.8 &     0.9  \\
S700B &   700.0 &  1384.1 &     1.0 &    55.2 &     0.9 &   615.8 &  4135 & 8246.1 &    54.6 &   376.9 &     0.9 \\
S1000A &  1000.0 &  1166.2 &     0.9 &    55.3 &     1.1 &   868.0 & 3908 & 11041.9 &    54.4 &   492.2 &     0.6 \\
S2000A &  2000.0 &  1115.5 &     1.0 &    55.7 &     0.5 &  1758.2 & 5353 & 10291.0 &    54.7 &  1250.0 &     0.3  \\
\enddata
\tablecomments{Each column has the following meaning:
$M_\mathrm{i}$: initial mass, 
$v_\mathrm{i}$: initial rotational velocity at the equatorial surface, 
$v_\mathrm{i}/v_\mathrm{c}$:  ratio of the initial rotational velocity to the critical velocity,  
$\log J_\mathrm{i}$: initial total angular momentum, 
$T_\mathrm{c, f}$: central temperature at the end of calculation,  
$M_\mathrm{f}$: final mass, 
$T_\mathrm{eff, f}$: final effective temperature, 
$R_\mathrm{f}$: final radius, 
$\log J_\mathrm{f}$: final total angular momentum, 
$M_\mathrm{He}$: mass of the helium core (i.e., the total mass below the hydrogen envelope) at the end of the calculation, 
$a_\mathrm{He}$: Kerr parameter  of the black hole that the entire helium core would make as a result of collapse. 
}
\end{deluxetable}

We calculated evolutionary models for 300, 500, 700, 1000, and 2000~\Msun{}
including rotation but without the TS dynamo as summarized in Table.~2.  We
exceptively included the TS dynamo in the sequence S300M that shares the same
initial condition with S300C, for comparison.  It is well known that these very
massive stars undergo the pair instability when the central temperature exceeds
about $10^9$~K, which leads to rapid contraction of the core and the consequent
rapid oxygen and silicon burning on a dynamical timescale.  Because of the very
high binding energy, this explosive nuclear burning cannot make these stars
explode to produce a supernova.  This pair instability phase was followed until
$T_\mathrm{c} =  5.4\times 10^{9}$~K and  $T_\mathrm{c} = 6.9\times 10^{9}$~K
for S300C and S500C respectively, which are good candidates for super-collapsar
progenitors.  Although these end points still did not reach the pre-collapse
stage ($T_\mathrm{c} \sim 10^{10}~\mathrm{K}$), we did not continue the
calculation because of a numerical difficulty  mainly caused by very small
timesteps encountered at such high temperature at the center.   For the other
model sequences, the calculation was terminated immediately before/at the onset
of the pair instability ($T_\mathrm{c} = 5\times10^8 \cdots 10^9$~K), beyond
which the evolutionary time is too short to have any further significant
redistribution of angular momentum. We find that the chemically homogeneous
evolution (CHE) does not occur in any of our model sequences (see YDL12 for a
detailed discussion on the condition of the CHE).

\begin{figure*}
\epsscale{1.00}
\plotone{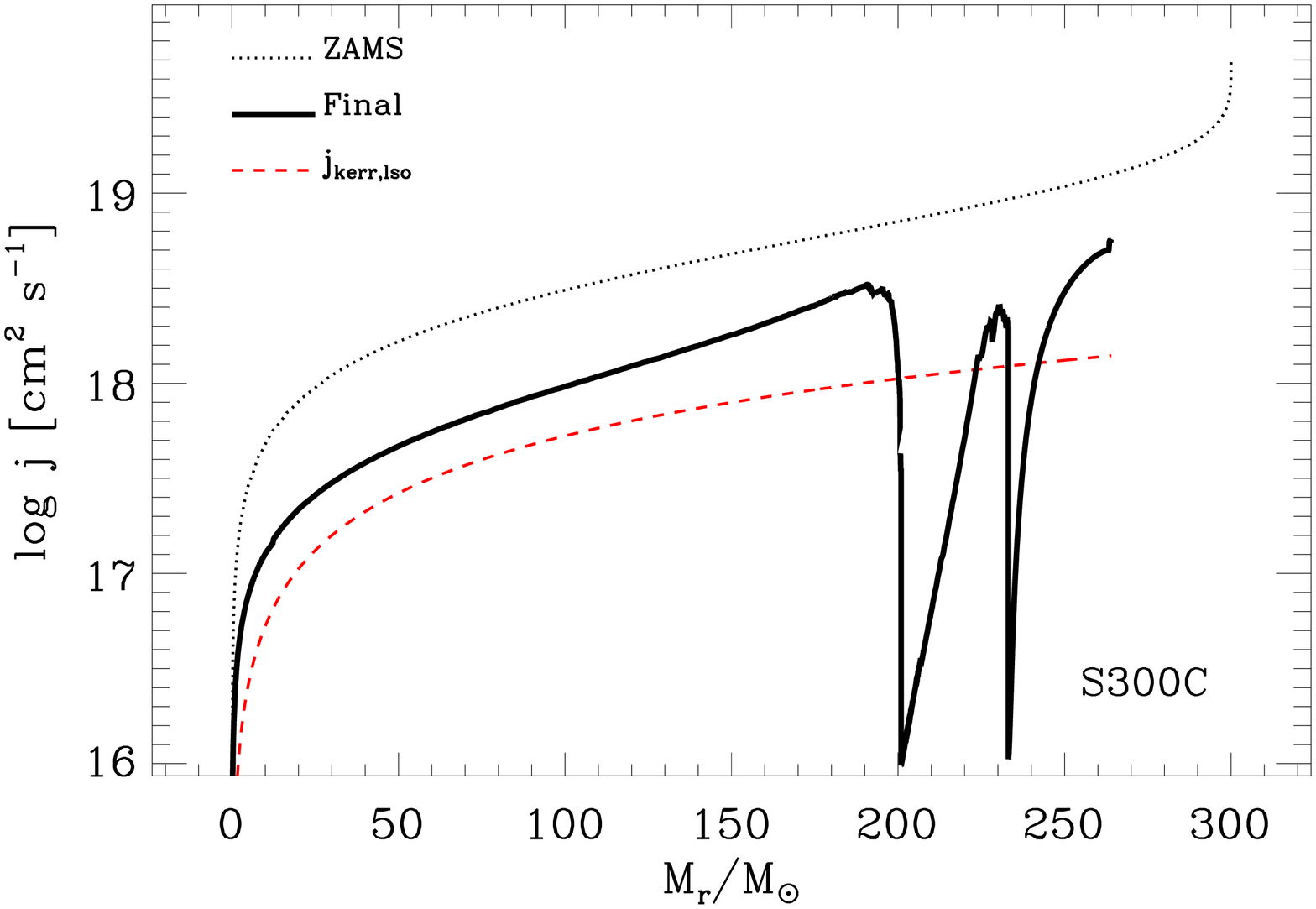}
\plotone{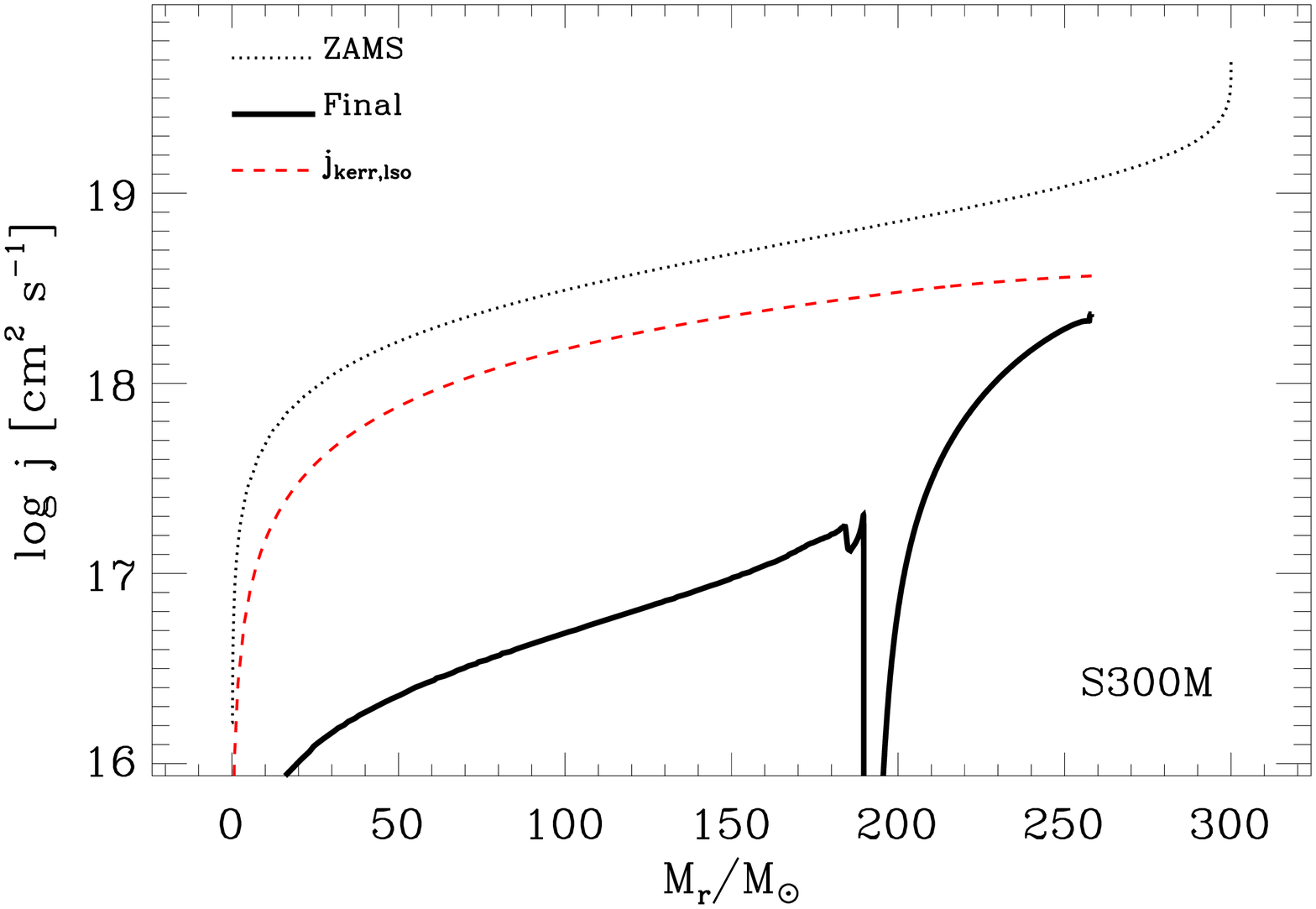}
\plotone{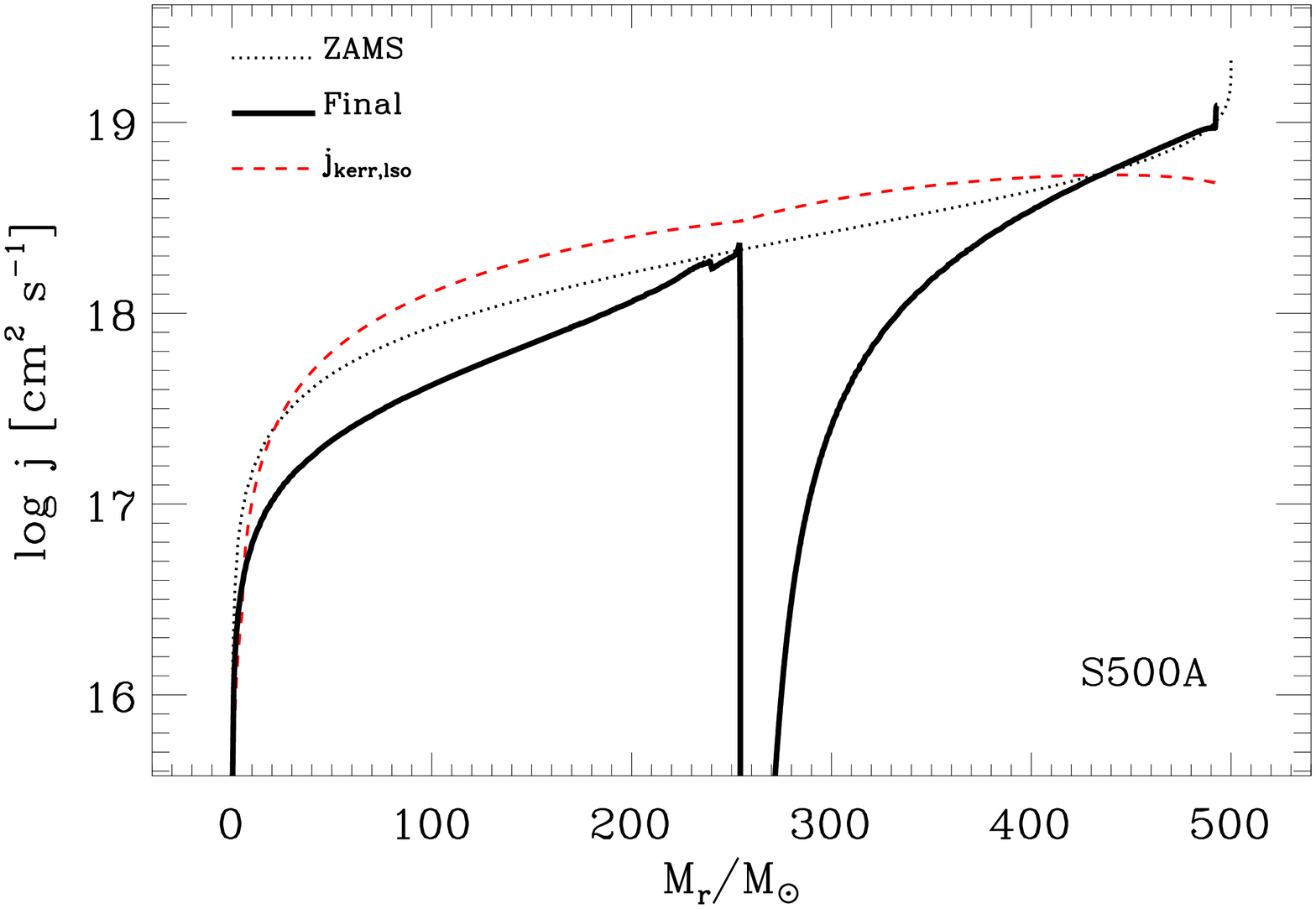}
\plotone{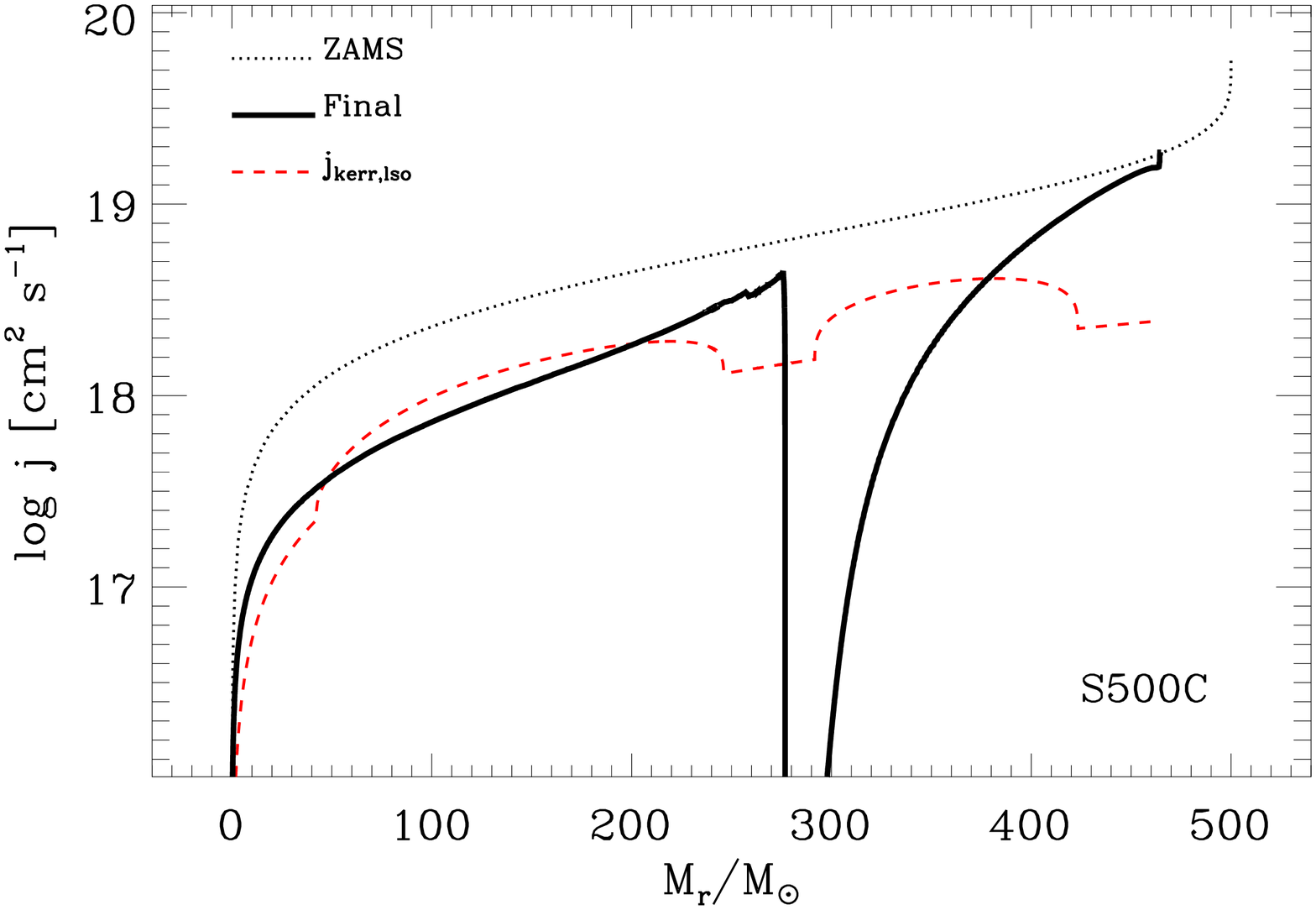}
\plotone{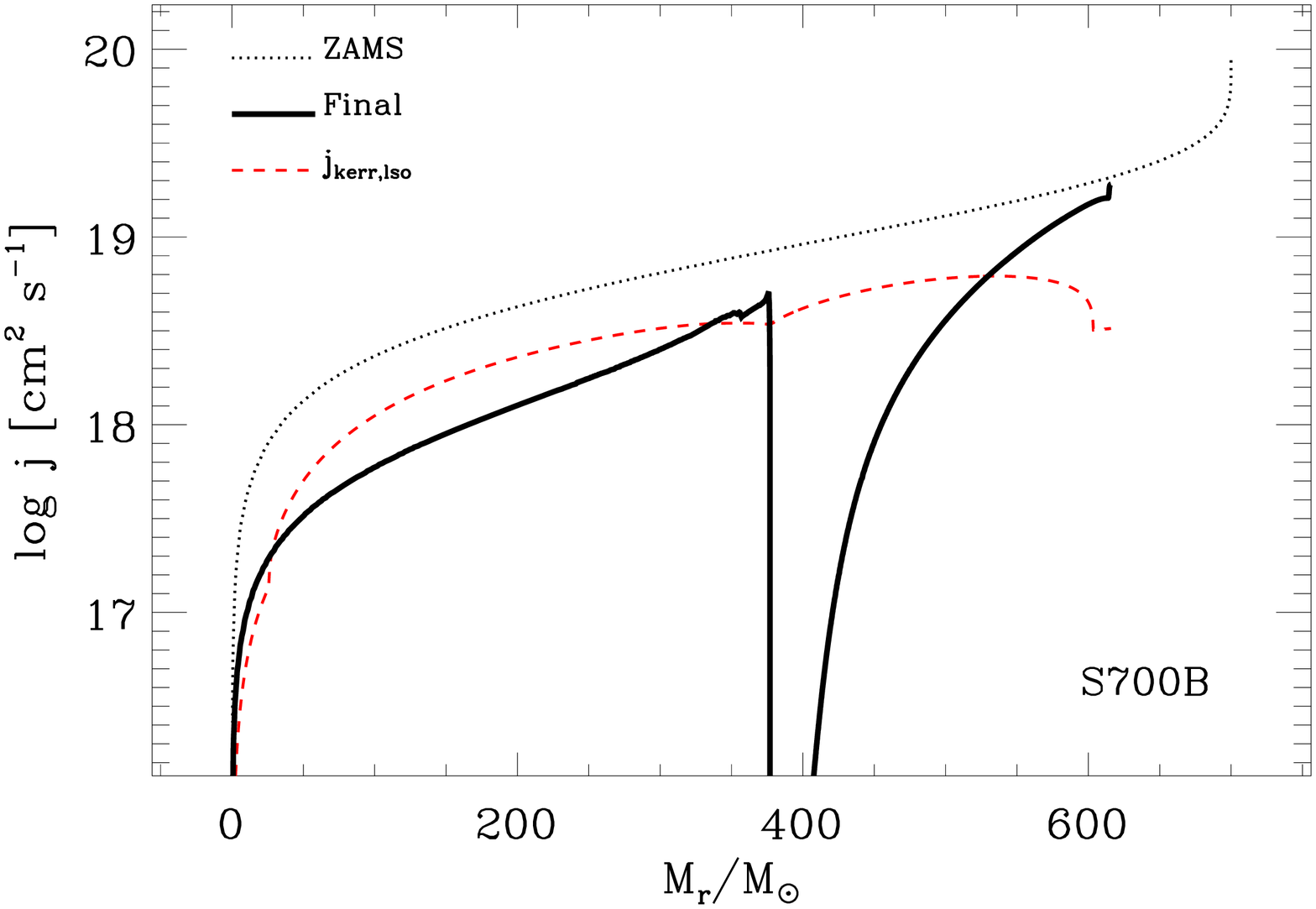}
\plotone{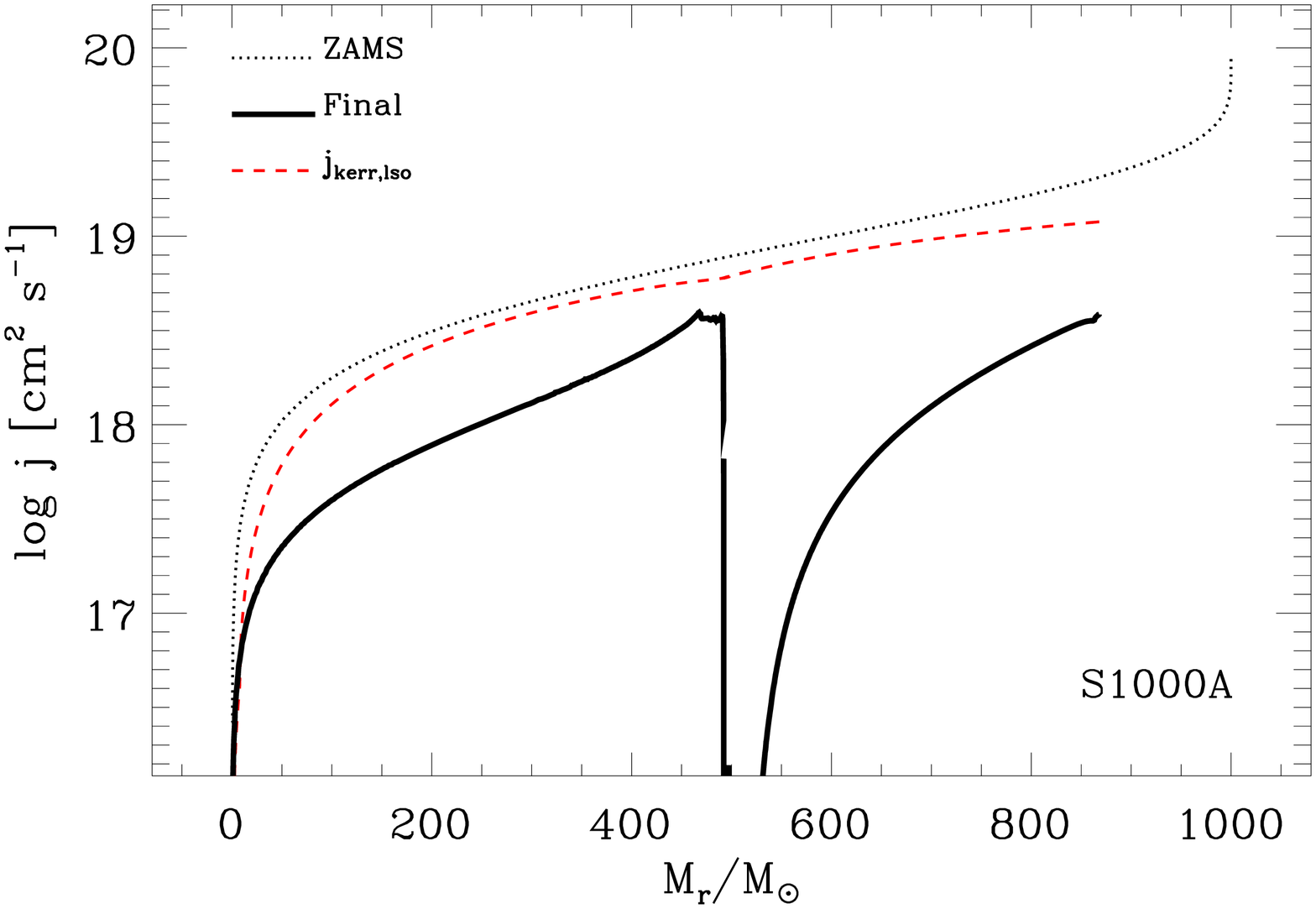}
\caption{Mean specific angular momentum over the shells. The initial and final distributions of specific angular momentum are given by the dotted 
and solid lines, respectively. The dashed line denotes the specific angular momentum for the last stable orbit around a Kerr black hole having 
all mass and angular momentum below the given mass coordinate.  
}\label{fig:jspec}
\end{figure*}

As shown in Fig.~\ref{fig:jspec}, most of the ZAMS models of the considered
mass range have  a sufficient amount of angular momentum  to produce a
collapsar, such that any layer of them could form an accretion disk around the
black hole that would be made below it.  As mentioned above, these stars on the
ZAMS rotate as a solid body.  As hydrogen burning in the core develops,
differential rotation across the boundary between the contracting convective
core and the expanding radiative envelope would be created without any
transport of angular momentum.  However, they undergo rapid redistribution of
angular momentum mostly via convection and Eddington-Sweet circulations as long
as the chemical stratification inside stars is not significant.  Convection
leads to rigid rotation in the convective core on the dynamical timescale, and
Eddington-Sweet circulations in chemically homogeneous layers occur on the
thermal timescale that is much shorter than the evolutionary
timescale\footnote{We find that the role of the secular shear instability and
the GSF instability is minor in our models.}. 

\begin{figure*}
\epsscale{1.00}
\plotone{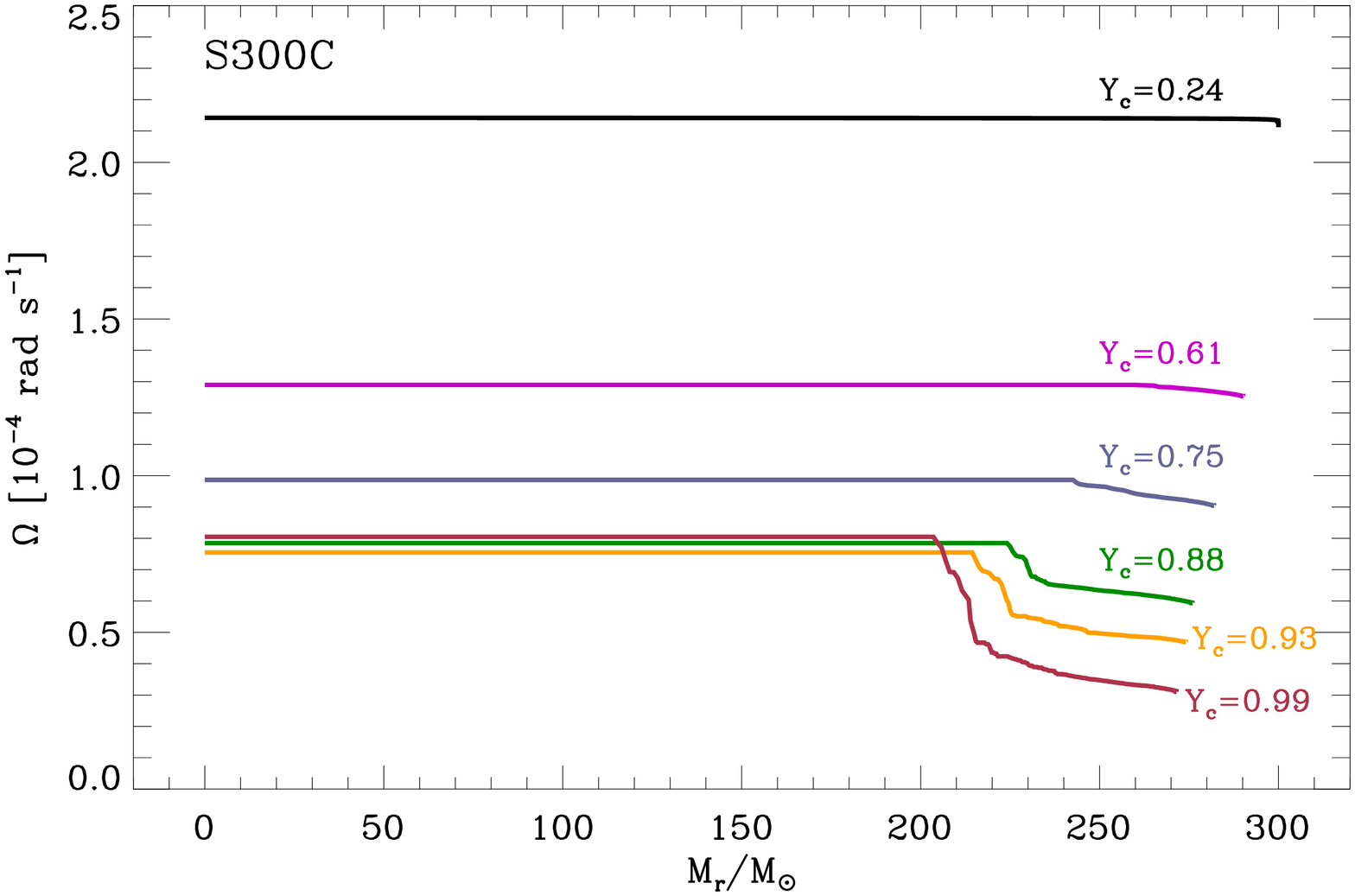}
\plotone{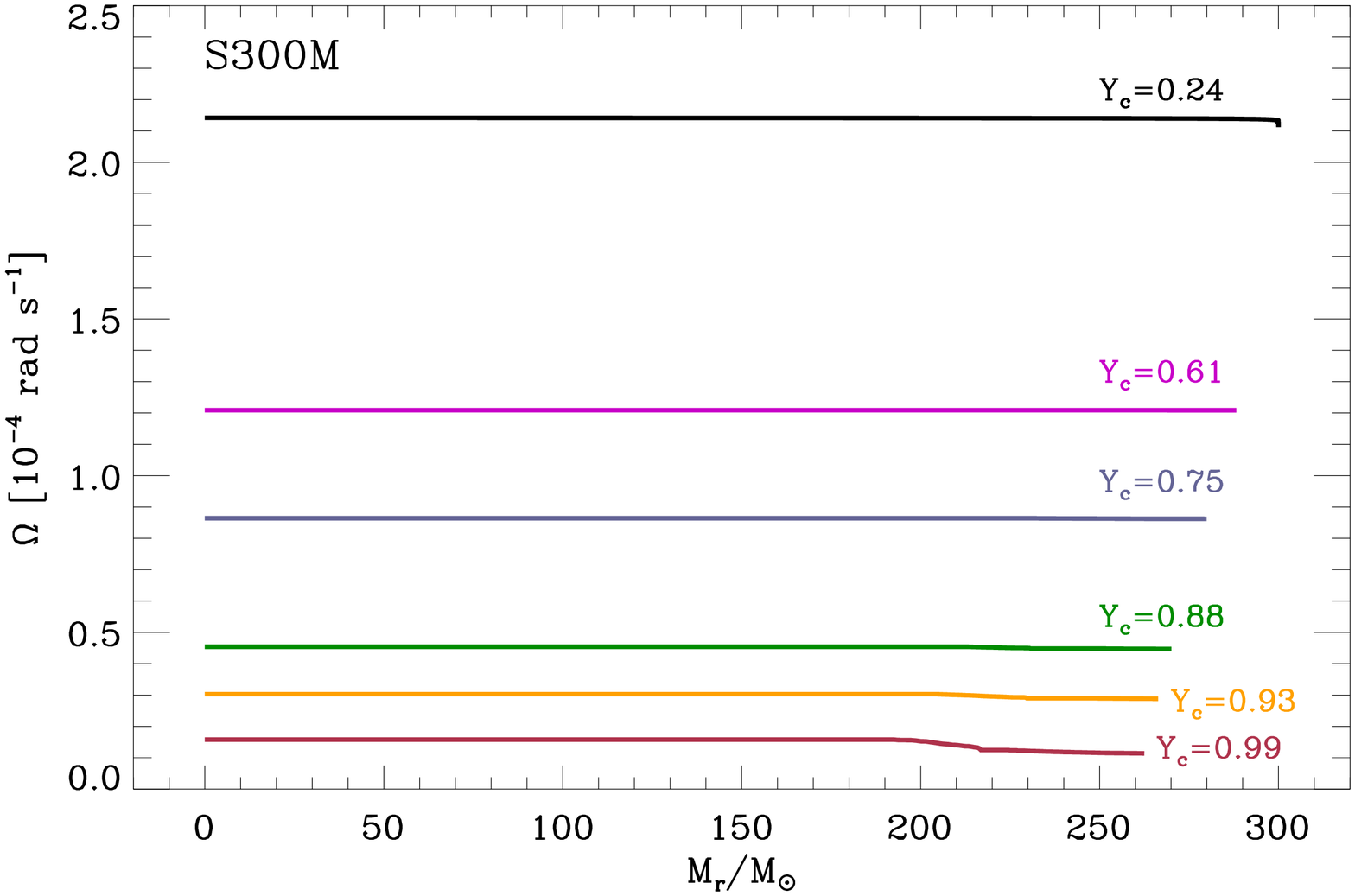}
\plotone{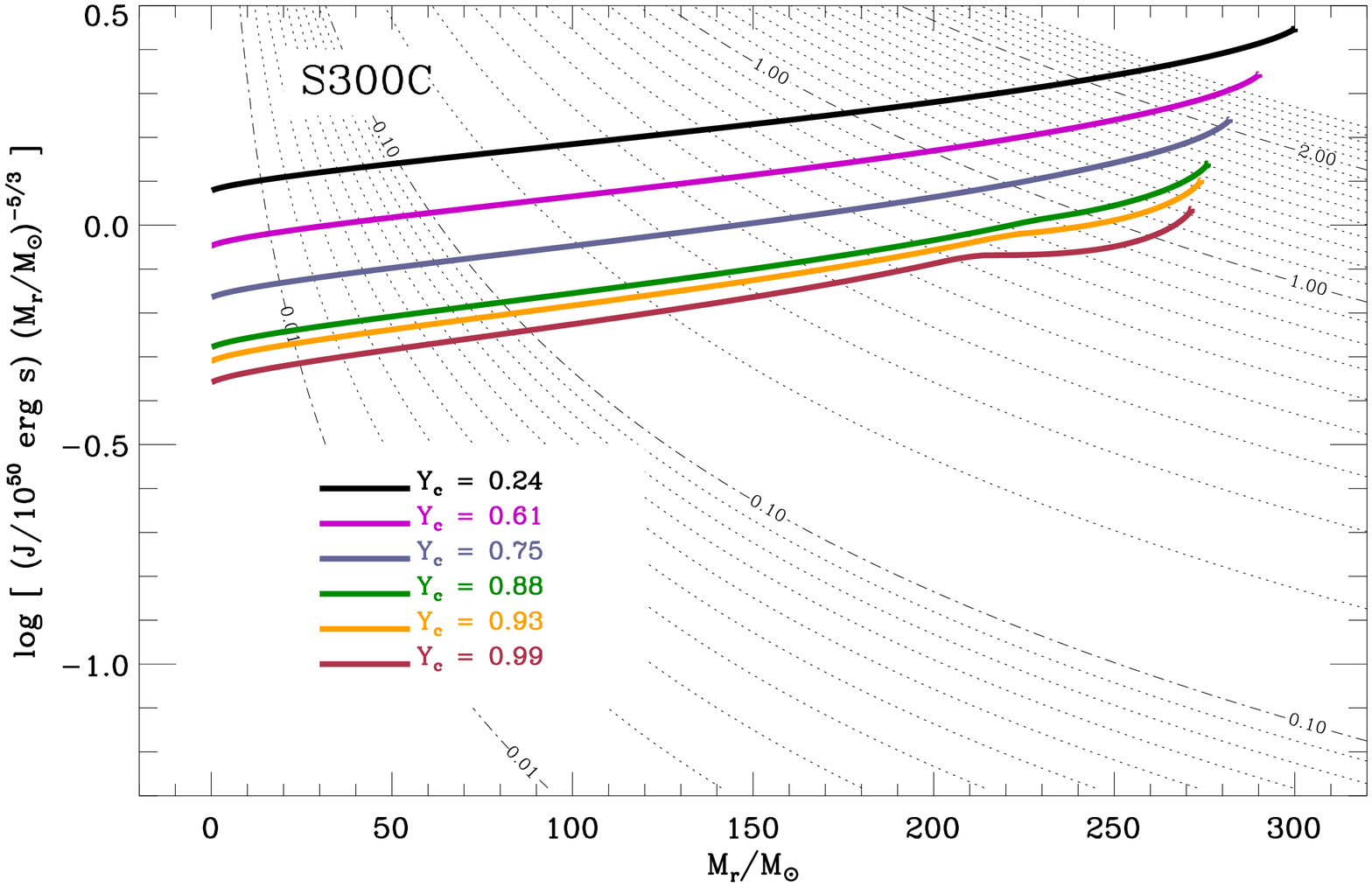}
\plotone{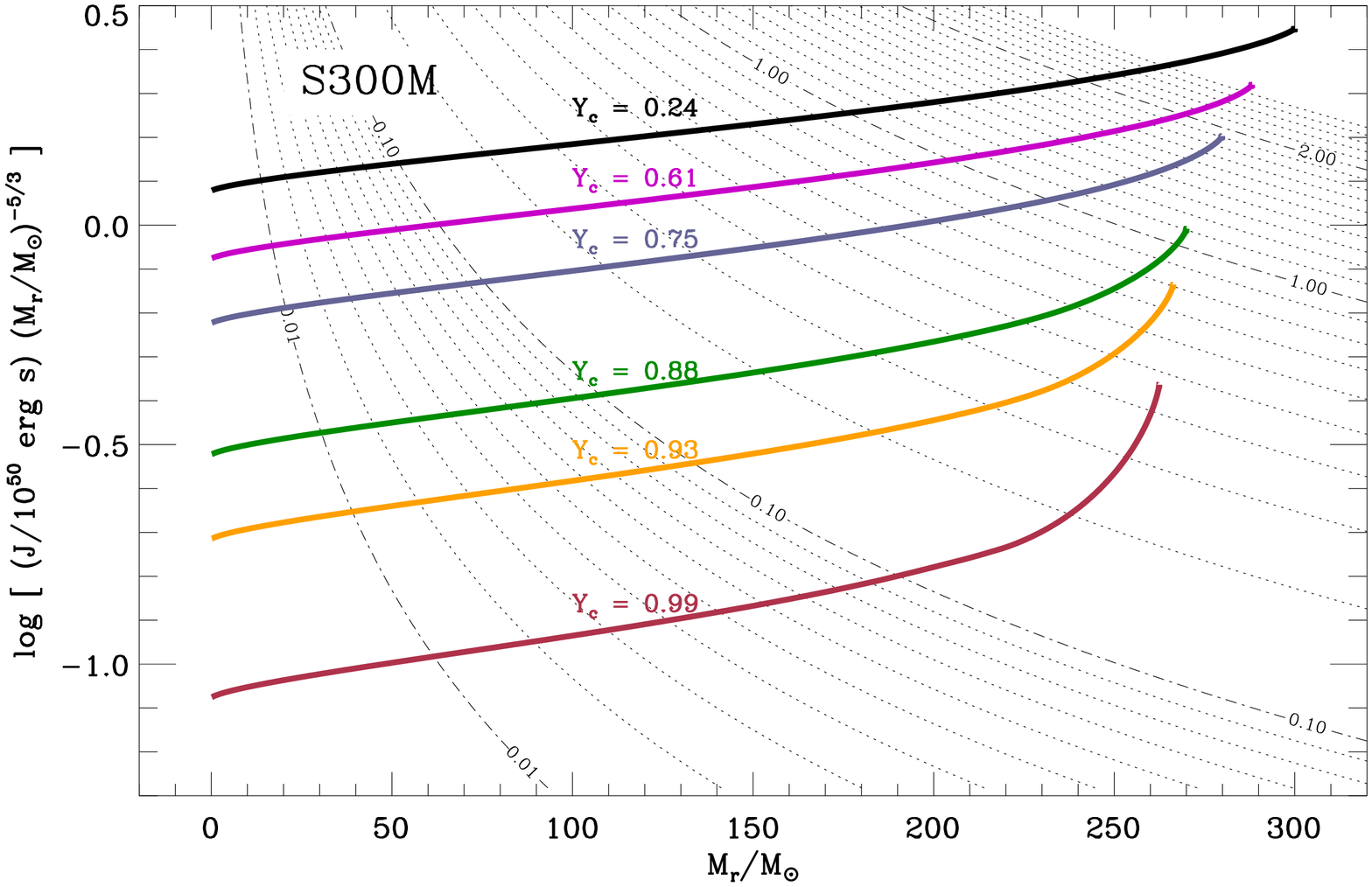}
\caption{\emph{Top panels}: Evolution of angular velocity on the main sequence in S300C (left panel) 
and S300M (right panel), respectively. The label $Y_\mathrm{c}$ marks the central mass fraction 
at the given evolutionary time of each profile.  
\emph{Bottom panels}: The corresponding  integrated angular momentum ($J(M_r)=\int_0^{M_r} j(m)dm$) profiles
divided by $M_r^{5/3}$. (The evolution of the angular momentum inside stars is better visualized with $J(M_r)/M_r^{5/3}$ 
than with $J(M_r)$ as discussed by \citet{Heger00}.) The thin contour lines denote levels of constant angular momentum in units of 
$10^{50}~\mathrm{erg~s}$. 
}\label{fig:omega}
\end{figure*}

For example, in the sequence S300C, solid-body rotation is maintained fairly
well until the central helium mass fraction reaches about 0.75 (The top-left
panel of Fig.~\ref{fig:omega}).  The star reaches the critical rotation soon
after the onset of core hydrogen burning because of this rapid transfer of
angular momentum and the gradual increase in the radius and surface luminosity,
and therefore loses angular momentum by the resulting mass loss.  However, the
angular momentum transfer from the core to the envelope via Eddington-Sweet
circulations is significantly slowed down when the chemical stratification
becomes strong across the boundary between the core and the envelope (the
so-called $\mu-$barrier; \citealt{Meynet97}).  Angular momentum  can be
effectively trapped in the core in this way and the degree of differential
rotation between the core and the envelope becomes gradually stronger
(Fig.~\ref{fig:omega}).  The final model of S300C retains enough angular
momentum to produce a collapsar (Fig.~\ref{fig:jspec}). 

In the corresponding magnetic model sequence S300M,  magnetic torques via the
TS dynamo are strong enough to overcome the $\mu-$barrier and near-rigid
rotation is maintained until the end of the main sequence (Top-right panel of
Fig.~\ref{fig:omega}).  As a natural consequence, this star loses 
more angular momentum than in S300C (Bottom panels in Fig.~\ref{fig:omega}).
In contrary to the corresponding non-magnetic model, the final model of S300M
does not retain enough angular momentum for a collapsar to occur (see YDL12 for
more comprehensive discussion on magnetic Pop III star models). 

The initial mass is another key factor to determine the final outcome.  The
envelopes of more massive stars expand more rapidly both on the main sequence
and during the post-main sequence phases \citep{Marigo03, Yoon12} and the
amount of angular momentum carried away by mass loss, which is proportional to
$R^2$, becomes larger.  In consequence, Fig.~\ref{fig:jspec} indicates that the
conditions for collapsar at the final stage can be fulfilled only for
$M_\mathrm{i} \lesssim 700$~\Msun{}. From this result, we conclude that the
initial mass range for potential super-collapsar progenitors is
$300~\mathrm{M_\odot} \lesssim M_\mathrm{i} \lesssim 700~\mathrm{M_\odot}$. 

\section{Implications for supernovae and gamma-ray bursts from super-collapsars}\label{sect:implications}

\begin{figure}
\epsscale{1.00}
\plotone{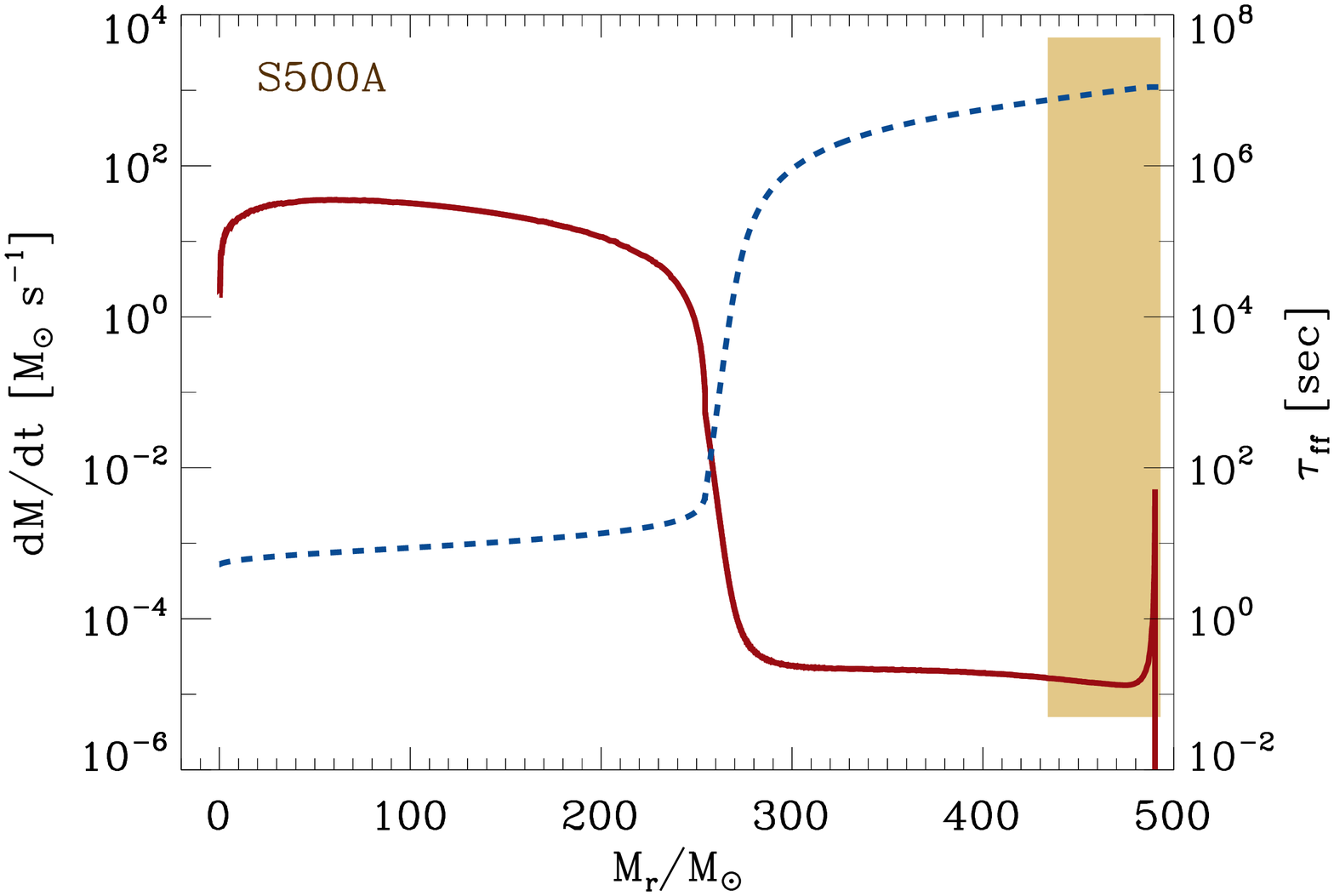}
\plotone{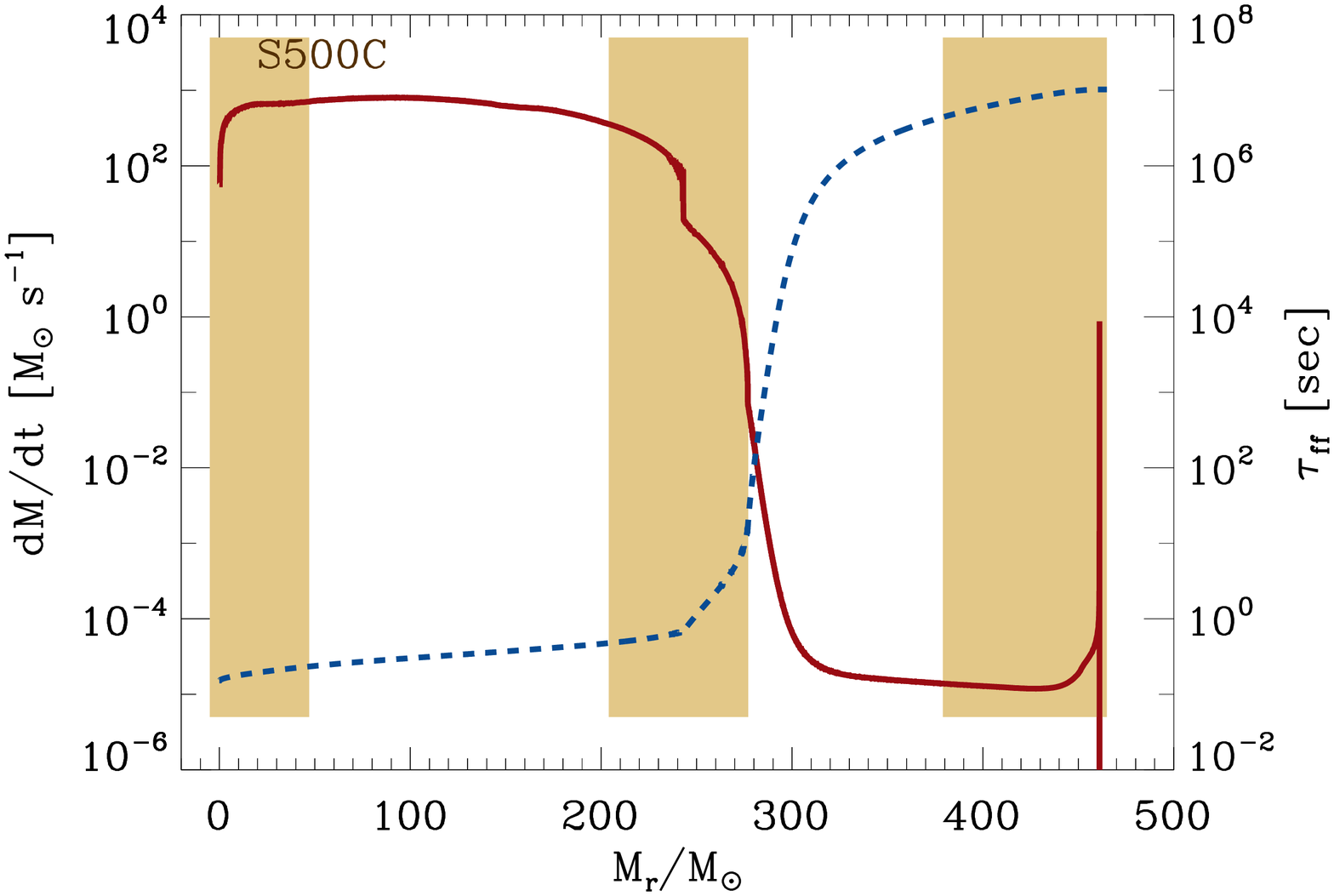}
\caption{The expected mass accretion rate onto the black hole (solid line) given by Eq.~(\ref{eq3}) as a function of the mass coordinate in the last models of S500A (upper panel) and S500C (bottom panel). The dashed line denotes the corresponding free-fall timescale ($\tau_\mathrm{ff} = 1/\sqrt{24\pi G\bar{\rho}}$, see the text). 
The rapidly rotating layers for which the specific angular momentum is higher than the critical value
for the formation of a Keplerian disk (see Fig.~\ref{fig:jspec}) are marked by the color shading. 
}\label{fig:grb}
\end{figure}

All of our very massive Pop III star models become a red supergiant (RSG; Table~2). 
It is well known that a jet produced in the stellar core via a collapsar cannot easily penetrate 
a RSG envelope and therefore our models may not be good 
progenitors of gamma-ray bursts, unfortunately. 
To investigate the possibility of a gamma-ray burst from our super-collapsar
progenitors in more detail, we made an estimate on the accretion rate from the disk
that the rapidly rotating layers would make around the black hole,  for 
the last models of S500A and S500C  as shown in
Fig.~\ref{fig:grb}. Here we followed the approximation by \cite{Woosley12}:  
\begin{equation}\label{eq3}
\dot{M} = \frac{2M_r}{\tau_\mathrm{ff}} \left( \frac{\rho}{\bar{\rho} - \rho}\right)~~, 
\end{equation}
where $\bar{\rho} = 3M_r/(4\pi r^3)$ and $\tau_\mathrm{ff} = 1/\sqrt{24\pi G\bar{\rho}}$. 
Although the last model of S500C is fairly close to the pre-collapse stage, 
the density of the core would further increase until the formation of the black hole, 
and  the accretion rate from the core material given in the figure should be considered a lower limit. 
The sequence S500A was terminated at the onset of carbon burning (see Table~2), which is still far from the pre-collapse stage, but 
the core does not retain enough angular momentum for collapsar in this case. 
The accretion rate from the envelope in both S500A and S500C should not be affected by this uncertainty, because
the hydrogen envelope has a much larger dynamical timescale than that of the core as shown in the figure. 
 
The figure indicates that  
there would be  three phases of mass accretion via an accretion disk in 
S500C.   The first two  would be made by accretion from the infalling matters of
$3~\mathrm{M_\odot} \lesssim M_r \le 47~\mathrm{M_\odot}$ (assuming the minimum
black hole mass of 3~\Msun{}) and of $204~\mathrm{M_\odot} \le M_r \le
277~\mathrm{M_\odot}$, and the last one from that of the hydrogen envelope
($379~\mathrm{M_\odot} \le M_r \le 465~\mathrm{M_\odot}$).  The expected
accretion rate in the core is very high, implying a production of a very
powerful jet.  However, the lifetime of the engine powered by the core
material would be very short.  Most of the core material below $M_r =
277~\mathrm{M_\odot}$  would be accreted onto the black hole within a second.
The outermost layer of the core around $M_r =  270 - 277$~\Msun{} has a
relatively low accretion rate, but the accretion time may not exceed 100
seconds.  This is much shorter than the jet crossing time through the hydrogen
envelope ($\tau_\mathrm{cross} \sim R_*/c \sim 10^4$~sec.), posing a serious
obstacle for the jet penetration and a gamma-ray burst (GRB) is not expected from the jet
produced by the core. 

The most interesting question may be whether or not a GRB-like event can be
made by the disk accretion of the matter in the layers of $434~\mathrm{M_\odot}
\le M_r \le 493~\mathrm{M_\odot}$ and $379~\mathrm{M_\odot} \le M_r \le
465~\mathrm{M_\odot}$ of the hydrogen envelopes in S500A and S500C,
respectively.  In particular, the expected accretion rate rapidly increases
near the surface in Fig.~\ref{fig:grb}, which results from the density
inversion in this layer (see \citealt{Woosley12} for more discussion on the
effect of the density inversion).  For the case of S500C,  a GRB-like event
is not likely to occur.  The energy of the jets from the core injected into the
hydrogen envelope would be higher than $10^{53}$~erg, if we assume the
accretion-to-jet conversion efficiency of about $\eta = 10^{-3}$ following
\citet{Suwa11}. Given that $\tau_\mathrm{cross}  << \tau_\mathrm{ff}$, the
forward shock of the jet from the core would become spherical by the time of
its breakout from the envelope.  Because the binding energy of the hydrogen
envelope is only about $10^{51}$~erg, the whole hydrogen envelope would be
blown away by the shock heating,  preventing accretion of the outer envelope
matter onto the black hole. The final outcome would be a jet-driven
supernova of Type IIP, rather than a GRB.  

By contrast, for the case of S500A, the inner layers are not rapidly rotating,
and  expected to directly collapse to a black hole without making a jet.  The
outermost layers of  the hydrogen envelope around the equatorial plane could be
accreted onto the black hole via an accretion disk to produce a GRB-like event
about two weeks after the collapse.  In this case,  with $\eta = 10^{-3}$, the
gamma-ray transient would have $L \sim 10^{47}~\mathrm{erg~s^{-1}}$ lasting for
about a month.  During the last few seconds, the accretion rate would
dramatically increase as shown in Fig.~\ref{fig:grb}, resulting in $L\sim
10^{49}~\mathrm{erg~s^{-1}}$. 

The above discussion is based on the S500A and S500C model, but the overall
conclusion would be the same with the other progenitor models: A GRB from very
massive Pop III stars is generally very difficult to occur because of the red
supergiant envelope.  Collapsar events in the inner layers would produce a
jet-powered type IIP supernova, if the initial rotation velocity of the
progenitor were close to the critical limit.  In a relatively slowly rotating
progenitor, the core would not retain enough angular momentum for collapsar,
but a gamma-ray transient could be produced with the rapidly rotating outermost
layers of the hydrogen envelope.  This event would be marked by a long-lasting
phase with a relatively low luminosity ($L \sim 10^{47} ~\mathrm{erg~s^{-1}}$
depending on $\eta$), which may resemble Swift 1644+57 \citep{Burrows11,
Levan11, Quataert12}, followed by a short phase of a few seconds having an
enhanced luminosity by factors of $10^2 - 10^4$ depending on the detailed
structure of the layer with density inversion, in the rest frame.  Such a
transient from redshift of $z\sim20$ would be bright in X-ray bands and  last
for a few years in the observer's frame.  Given that $L\gtrsim
10^{52}~\mathrm{erg~s^{-1}}$ is needed for a GRB from redshift of about 20 to
be detected by the X-ray detector BAT \citep{Komissarov10}, the predicted
gamma-ray luminosity is too low to be observed in near future, unfortunately. 

One uncertain factor in our models is the mass-loss rate from very massive Pop
III stars during the red supergiant phase.  If the whole hydrogen envelope were
stripped off before the collapse, favorable conditions for GRB production would
be more easily fulfilled for the progenitors with rapidly rotating cores.  As
explained above, we considered enhancement of mass loss due to surface
enrichment of heavy elements via chemical mixing by extrapolating the mass-loss
rate given by \citet{Nieuwenhuijzen90}.  In all of our models, the surface
metallicity remains smaller than about a few times $10^{-7}$, and the
consequent enhancement of the mass loss rate is not significant.  The mass loss
rate might be further influenced by pulsational instabilities that can easily
occur in stars close to the Eddington limit.  \citet{Baraffe01} concluded that
mass loss induced by pulsation would not be efficient: only about 24~\Msun{} is
expected to be lost from a 500~\Msun{} star.  This is not significant compared
to the amount of the lost mass in our models. However, the mechanism of RSG
winds and its dependence on the surface abundance of CNO elements are  not
well understood yet and remains the biggest uncertain factor in our conclusions.  

Another issue that is worth discussion is the final structure of very massive
stars. \citet{Suwa11} and \citet{Nagakura12} showed that the GRB jet can
breakout a very massive Pop III star if it is a blue supergiant (BSG)
progenitor, for which accretion from the long-lived  hydrogen envelope to the
black hole plays a key role \citep[see also][]{Woosley12}.  Some peculiar GRBs
like GRB 130925A may be well explained by a progenitor of this type as recently
argued by \citet{Piro14}.  The relevant question is whether or not very massive
Pop III stars can end its life as a BSG, rather than a RSG.  The BSG model of
915~\Msun{} used by \citet{Suwa11} and \citet{Nagakura12}  is taken from
\citet[][the Y-1 model]{Ohkubo09}, who constructed this model with rapid mass
accretion from a 1~\Msun{} protostar.  By contrast, in the present study as
well as in the previous ones by  \citet{Baraffe01}, \citet{Marigo03} and YDL12,
all Pop III stars with $M \gtrsim 300~\mathrm{M_\odot}$ become  RSGs at the
final stage for both non-rotating and rotating cases, unless they lose most of
their hydrogen envelopes.  This result seems robust given that each group used
a different prescription for overshooting, semi-convection and rotation.  To
further investigate  how the final radius varies with different assumptions on
the overshooting and semi-convection, we calculated a couple of models without
overshooting and a small semi-convection parameter ($\alpha_\mathrm{semi} =
0.04$) with 300~\Msun{} and 500~\Msun{} and compared them with our fiducial
models.  As presented in Table~3, this hardly causes any difference in the
final radius  for 300~\Msun{}.  A significant reduction in the final radius
with no-overshooting is found with 500~\Msun{} but it still does not become a
BSG. This strong tendency to become a RSG might be related to the fact
that very massive stars are very close to the Eddington limit: according to the
equation of state, more expansion of a gas is
needed to consume a given amount of energy input as
the role of radiation pressure becomes more important.

The discrepancy between \citet[][in particular the result with their Y
sequences]{Ohkubo09} and the others is probably related to the rapid mass
accretion throughout the whole evolutionary stages that was assumed in Ohkubo
et al.. This assumption is, however, subject to many uncertainties.  For
example, it is observed that in their M-1 sequence, for which mass accretion
was stopped for $M \ge 321~\mathrm{M_\odot}$,  the radius rapidly increases to
$\sim 1000~\mathrm{R_\odot}$ during the post-main sequence phase, unlike the
case  of the Y sequences where rapid mass accretion continues until the end
of the evolution.  Note that the M-1 sequence in Ohkubo et al. was constructed
to consider the radiation feedback from the mass accreting star that can
greatly reduce the mass accretion rate as discussed by \citet{McKee08}.  The
feedback effect must become more complicated with rotation. In particular, mass
accretion should result in accretion of angular momentum and the mass-accreting
Pop III proto-stars could easily reach the critical rotation when the mass
grows beyond a certain limit.  Given that mass accretion would be effectively
halted once the stellar surface reaches the critical rotation, the Ohkubo et
al.'s assumption of continuous mass accretion that gave the BSG solution  is
not likely to be valid with rotation.  This issue will be addressed in a
separate paper (H. Lee and S.-C.  Yoon, in preparation).

Our conclusion is still subject to modifications with binary interactions.  For
example, in a close binary system, the mass accreting star may not be easily
rejuvenated but have a relatively small core mass compared to the total mass,
if mass accretion occurs near the end of the main sequence or during the
post-main sequence phase.  This can often make the star remain blue during the
late evolutionary stages \citep{Podsiadlowski89, Braun95}. The evolution of
massive binary Pop III stars is thus an interesting topic of future work. 

\begin{deluxetable}{ccccccc}
\tabletypesize{\scriptsize}
\tablecaption{Final radius of very massive stars for different mixing parameters}\label{tab3}
\tablewidth{0pt}
\tablehead{
\colhead{$M_\mathrm{i}$} & 
\colhead{$v_\mathrm{crit}/v_\mathrm{K}$} & 
\colhead{$f_\mathrm{over}$} & 
\colhead{$\alpha_\mathrm{semi}$} & 
\colhead{$M_\mathrm{f}$} & 
\colhead{$T_\mathrm{eff, f}$} &  
\colhead{$R_\mathrm{f}$} 
}
\startdata
[$\mathrm{M_\odot}$] &   &   &    & [$\mathrm{M_\odot}$]  & [K] &  [$\mathrm{R_\odot}$] \\        
\hline
300 & 0.9     & 0.000 & 0.04  &  251.9  & 3821  & 4761.8  \\
300 & 0.9     & 0.335 & 1.00  &  264.0  & 4271  & 4769.8  \\
500 & 0.8     & 0.000 & 0.04  &  469.8  & 7293  & 2550.6  \\
500 & 0.8     & 0.335 & 1.00  &  464.9  & 4157  & 7022.7  \\
\enddata
\tablecomments{Each column has the following meaning:
$M_\mathrm{i}$: initial mass, 
$v_\mathrm{crit}/v_\mathrm{K}$: the critical rotation for the given $\Gamma$ at the equatorial surface in units of the Keplerian value with rigid rotation, 
$f_\mathrm{over}$: overshooting parameter. I.e.,  the convective layer is extended by $f_\mathrm{over}$ times 
local pressure scale heights  beyond the convectively unstable region, 
$\alpha_\mathrm{semi}$: semi-convection parameter as defined by \citet{Langer83}, 
$M_\mathrm{f}$: final mass, 
$T_\mathrm{eff, f}$: final effective temperature, 
$R_\mathrm{f}$: final radius. 
}
\end{deluxetable}

\section{Conclusions}\label{sect:conclusions}

We discussed the possibility of super-collapsars using the evolutionary models
of very massive Pop III stars.  We find that Pop III stars with $M_\mathrm{i} >
3000~\mathrm{M_\odot}$ cannot be born with enough angular momentum for
collapsars in the core. This is mainly because these very massive stars are
very close to the Eddington limit for which case the critical rotation velocity
at the equatorial surface is severely limited to a value significantly below
the Keplerian velocity. The potential progenitors for super-collapsars that can
lead to energetic GRBs should have the initial mass range of
$300~\mathrm{M_\odot} \lesssim  M_\mathrm{i} \lesssim 3000~\mathrm{M_\odot}$.  

We have to consider the evolution of these stars to get a more realistic mass
range for super-collapsar progenitors.  High angular momentum and strong
magnetic fields of large scales at the pre-collapse stage are the two essential
conditions needed for super-collapsars \citep{Komissarov10, Meszaros10}.  Given
that magnetic torques  can lead to efficient braking of the stellar core, these
two conditions may not be easily satisfied simultaneously \citep[cf.][]{Yoon12}.  Our
evolutionary models  indeed show that the angular momentum condition can be
fulfilled only when we ignore magnetic torques.  If we only consider
hydrodynamic processes for the transport of angular momentum  like
Eddington-Sweet circulations, the angular momentum conditions for
super-collapsars at the final evolutionary stage can be fulfilled  for the
initial mass range of $300~\mathrm{M_\odot} \lesssim M_\mathrm{i} \lesssim
700~\mathrm{M_\odot}$. 

However,  these stars become red supergiants at the pre-collapse
stage and therefore production of an energetic gamma-ray burst from super-collapsar events
seems difficult. If the core can retain enough angular momentum to produce
relativistic jets, it may lead to a jet-powered type IIP supernova.  
If the initial rotational velocity is relatively low for the considered mass
range, only the outermost layers of the hydrogen envelope  may have specific
angular momentum above the critical limit for collapsar.  The consequent jet
formation would produce an ultra-long (about a month), relatively faint ($L\sim
10^{47}~\mathrm{erg~s^{-1}}$) gamma-ray transient that is marked by a short
spike in the gamma-ray luminosity ($L\sim 10^{49} -
10^{51}~\mathrm{erg~s^{-1}}$) during the last few seconds, in the rest frame.  

\acknowledgments

This work was supported by Basic Science Research (2013R1A1A2061842) program through the National Research Foundation of Korea (NRF).

\end{document}